\documentclass[11pt,a4paper]{article}
\pdfoutput=1
\usepackage{jcappub}

\usepackage{bm}
\usepackage{amsfonts}
\usepackage{placeins}

\newcommand{\be}{\begin{equation}}
\newcommand{\ee}{\end{equation}}
\newcommand{\bea}{\begin{eqnarray}}
\newcommand{\eea}{\end{eqnarray}}

\newcommand{\fabs}[1]{\left| #1 \right|}

\newcommand{\di}[1]{\text{#1}}
\renewcommand{\vec}[1]{{\bm #1}}
\begin{document}



\title{On the local variation of the Hubble constant}

\author[a]{Io Odderskov}
\author[a,b]{Steen Hannestad}
\author[c]{Troels Haugb{\o}lle}

\affiliation[a]{Department of Physics and Astronomy\\
 University of Aarhus, DK-8000 Aarhus C, Denmark}

\affiliation[c]{Aarhus Institute of Advanced Studies,\\
 Aarhus University, DK-8000 Aarhus C, Denmark}

\affiliation[c]{Centre for Star and Planet Formation\\
                Natural History Museum of Denmark and Niels Bohr Institute\\
                University of Copenhagen, DK-1350 Copenhagen, Denmark}

\emailAdd{isho07@phys.au.dk, sth@phys.au.dk, troels.haugboelle@snm.ku.dk}

\abstract{We have carefully studied how local measurements of the Hubble constant, $H_0$, can be influenced by a variety of different parameters related to survey depth, size, and fraction of the sky observed, as well as observer position in space. Our study is based on N-body simulations of structure in the standard $\Lambda$CDM model and our conclusion is that the expected variance in measurements of $H_0$ is far too small to explain the current discrepancy between the low value of $H_0$ inferred from measurements of the cosmic microwave background (CMB) by the Planck collaboration and the value measured directly in the local universe by use of Type Ia supernovae. This conclusion is very robust and does not change with different assumptions about effective sky coverage and depth of the survey or observer position in space.}
\maketitle

\section{Introduction}

Over the past decade the cosmological standard model based on the $\Lambda$CDM paradigm has been firmly established and tested using a variety of data.
One of the fundamental parameters of the $\Lambda$CDM model is the present expansion rate of the universe, $H_0$, also known as the Hubble constant.
The value of this parameter can either be inferred indirectly from e.g.\ measurements of the cosmic microwave background at high redshift or from large scale 
structure surveys at intermediate or low redshift, or it can be measured directly by measuring velocities and distances to standard candles in the nearby universe.

Both methods have been used successfully in the past, and the precision of both methods is now at a level where the agreement between the two types of measurements is a constraining consistency test of the underlying model. Both methods have advantages and disadvantages: The direct measurement is very model independent, but prone to systematics related to local flows and the standard candle assumption. On the other hand, the indirect method is very robust and precise, but relies completely on the underlying model to be correct.
Any disagreement between the two types of measurements could in principle point to a problem with the underlying $\Lambda$CDM model.

The recently released data from the Planck satellite seems to point to a relatively low value of the Hubble constant, while direct measurements relying mainly on Type Ia supernovae clearly favour a higher value, with the two being inconsistent at the $2.4\sigma$ level \cite{Riess:2011, Planck:2013}.

There are several possible explanations for the disagreement between the two estimates of the expansion rate. It could be caused by a problem with the assumed cosmological model, or it could be that one (or both) of the estimates is either inaccurate or biased. The second possibility has recently been considered by Estathiou in \cite{Efstathiou:2013} and by Clarkson et al. in \cite{Clarkson:2014}. Using an improved distance calibration, Estathiou has re-analysed the data from \cite{Riess:2011} and found a lower value for the local Hubble parameter, which decreases the tension between the two estimates. Ref.~\cite{Clarkson:2014} consider relativistic corrections to the distance to the CMB last scattering surface, and show that second order lensing corrections can possibly increase this distance by several percent, which in turn causes an increase in the best fit value of the Hubble parameter from the Planck data. 
Yet another possibility - the one that is tested in this paper - is that it is merely a result of the spatial variation in the expansion rate of our universe.\\
\\
\noindent
For a given cosmological model, the expansion rates estimated by observers at different locations are expected to vary according to some underlying distribution. To gain knowledge of the spread of this distribution, we perform N-body simulations of the model in question. Subsequently, structures are identified using a halo finder that generates a catalogue containing the masses, positions, velocities and substructures of all halos found in each simulation. From the halo catalogues, lists of observers are selected, under the assumption that the standard observer reside in a halo similar to the Local Group. Each observer is assumed to estimate the local Hubble parameter in his neighbourhood by measuring the distances and velocities of halos located in a sphere around him, since this is where we would expect supernovae to be found. Finally, confidence intervals are constructed from the local Hubble parameter calculated by each of these observers. This is done in order to answer the question: Could the discrepancy between the inferred and the measured Hubble constants be due to cosmic variance?

The same question has been addressed in \cite{Marra:2013} and \cite{Wojtak:2013}. In \cite{Marra:2013}, Marra et al. calculate the variation of local Hubble parameters based on the power spectrum which expresses the variations of the underlying density field, and this in turn affects the velocity field and hence the local Hubble parameters. In \cite{Wojtak:2013}, Wojtak et al. use the same approach as in this paper, i.e. N-body simulations, to estimate the spread in the local Hubble parameters. In their paper, they focus primarily on how the local Hubble parameters are affected by observer positions. While this is also discussed in the present paper, we furthermore investigate how the apparent variation is affected by only observing a fraction of the sky, and whether the cosmic evolution that takes place between the time that light is emitted by a supernova and observed by us can have a significant effect.

We finally note that measurements of variations in the Hubble parameter using type Ia supernovae is equivalent to measurements of the local velocity field using the same tracers (see e.g.\ \cite{Haugboelle:2006uc,Ma:2013oja,Feindt:2013pma,Turnbull:2011ty,Davis:2010jq,Weyant:2011hs}).
\\
\noindent
The paper is organized as follows: In section 2, we describe the N-body-simulations that our analysis is based upon and the halo finder used to identify structure. In section 3, the calculation of the local Hubble parameters is outlined, and the selection of observers and of observed halos is motivated and described. In section 4, we present the outcome of the analysis and finally we discuss the results in section 5. 

\section{Method}

Below the N-body simulations and identification of halos are described in detail. 

\subsection{Simulations}
The best fit cosmological parameters determined by the Planck collaboration, as specified in the last column of table 5 in  \cite{Planck:2013}, are used as the basis for N-body simulations of the universe. The N-body simulations are performed using a modified version of the GADGET-2 code \cite{Springel:2005}, with initial conditions generated using a code written by J.~Brandbyge \cite{Brandbyge:2006} based on transfer functions computed using CAMB \cite{Lewis:2002}. Specifically, the transfer functions are calculated with $(\Omega_b,\Omega_{CDM})=(0.048,0.26)$, whereas only cold dark matter is used in the N-body simulations. A flat universe is assumed, and $(h,\sigma_8)=(0.68,0.84)$. The simulations are run from a redshift of $z=50$ until $z=0$.

The reference simulation is done in a box of sidelength $512 \di{Mpc/h}$ with $512^3$ particles. We perform two simulations of this size and resolution, but with different seeds for the random number generator that is used to construct the initial conditions, in order to determine how much the results can be expected to vary between different occurrences of a universe with the same parameters. In order to check for numerical effects due to the simulation resolution and the simulation volume, we also perform two simulations with $1024^3$ dark matter particles in boxes of $512 \di{Mpc/h}$ and $1024 \di{Mpc/h}$, respectively. By comparing the results obtained from the standard simulation with the results obtained using a bigger box and a greater resolution, we conclude that this does not significantly affect the results, and that the chosen box size and resolution are sufficient to make the desired inquiries. Still, all the simulations are of much smaller volume than the ones that are the basis for the analysis performed by Wojtak et al. in \cite{Wojtak:2013}, which are of several $\di{Gpc/h}$, while the mass resolution is similar. The sample on which they base their analysis of the Hubble flow is therefore substantially larger than the ones used in this paper, and as a result the analyses presented here will have a greater sample noise and a greater correlation between the observers. The latter is to some extend mitigated by using several boxes with independent random seeds. We identify 600 observers in each simulation, and use these to estimate of the mean and spread of the Hubble constant.

The observers in the simulations are spread across the entire simulation volume. In order for each of them to be able to observe in all directions, the box is taken to be periodic. To avoid an observer considering the same halo twice in his calculation of the Hubble constant, the maximal observed distance should be half a box length, that is $256\di{Mpc/h}$, corresponding to a redshift of $z = 0.087$. This corresponds approximately to the greatest observed distance of $z \sim 0.1$ used in supernova surveys, \cite{Riess:2011}. As in \cite{Riess:2011}, in order to reduce the effect of the local, coherent flow, no halos closer than $30\di{Mpc/h}$, corresponding to $z = 0.01$, are used.

\subsection{Identifying halos}

To identify bound structures in the simulations described above, we make use of the halo finder ROCKSTAR \cite{Behroozi:2011,Manual:Rockstar} which identifies halos using a variant of the Friends-of-Friends (FOF) algorithm. At first, particles that are close together in space are grouped together using the FOF algorithm, which assembles particles that are within a specified distance of each other. Within each FOF-group, a measure of the phase-space distance between two particles $p_1$ and $p_2$ (with positions $\vec{x_1}$, $\vec{x_2}$ and velocities $\vec{v_1}$, $\vec{v_2}$, respectively) is defined as $d(p_1,p_2) = \sqrt{\frac{\fabs{\vec{x_1}-\vec{x_2}}^2}{\sigma_x^2}+\frac{\fabs{\vec{v_1}-\vec{v_2}}^2}{\sigma_v^2}}$, where $\sigma_x^2$ and $\sigma_v^2$ are the variances of the particle positions and velocities within the group. This metric combines distance in position and velocity into a single measure and, using this, locations in which the mean phase-space-distance between particles is low are identified as local maxima of phase-space density. These maxima are used as seed halos, and all particles in the original FOF-group are assigned to the seed that is closest in phase-space. At last, halos located in more massive hosts are categorized as subhalos, unbound particles are removed, and halo masses are calculated as spherical overdensities. For each halo, the position of the halo centre is calculated as the mean of the positions of the central particles, and the halo velocity is calculated as the mean of the velocities of the innermost 10\% of the halo particles.

\section{Analysis}

As mentioned, it is assumed that each observer estimates the local Hubble parameter, $H_{loc}$, by measuring velocities and positions of halos in the close vicinity. The apparent velocities of observed halos contain two components: one from the expansion of space, and one from peculiar motion. Assuming a pure Hubble flow, corresponding to no peculiar motion, the radial velocities would be given by Hubble's law: 

\begin{align}
v = H_0 r,
\end{align}
where $r$ is the radial distance to the halo and $H_0$ is the global Hubble constant. Assuming that the observers cannot determine how large the peculiar component of the radial velocity is, they fit the distances and velocities of the halos around them to Hubble's law, using a least squares estimate, hereby getting a value for the local Hubble constant:

\begin{align}
H_{loc}=\frac{\bar{r}\bar{v}}{\bar{r^2}},
\end{align}
where a bar denotes the mean and $v$ is the absolute velocity of the halo away from the observer, meaning that each observer has transformed to the CMB rest frame.

\begin{figure}
\center
\includegraphics[width = 0.49\textwidth]{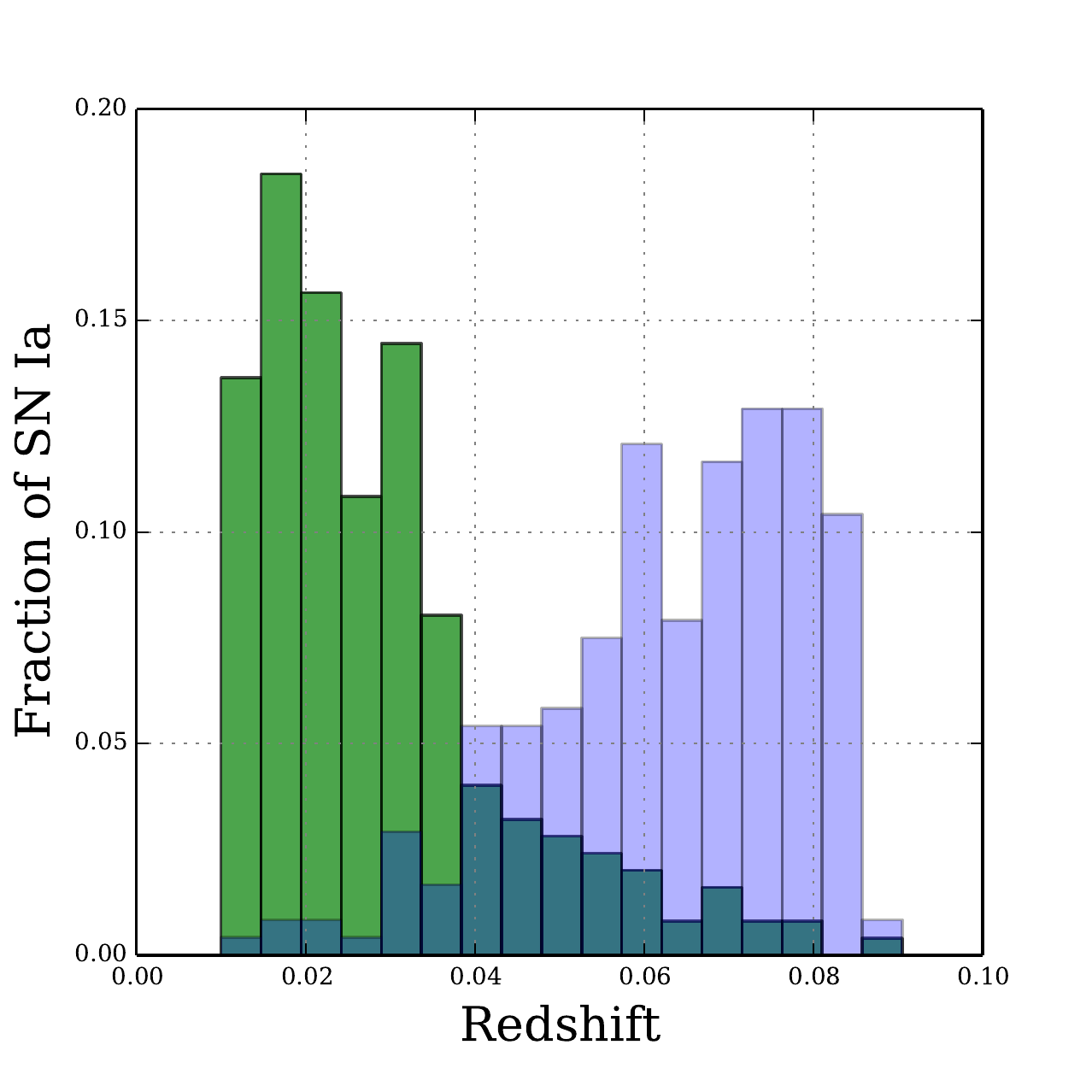} 
\includegraphics[width = 0.49\textwidth]{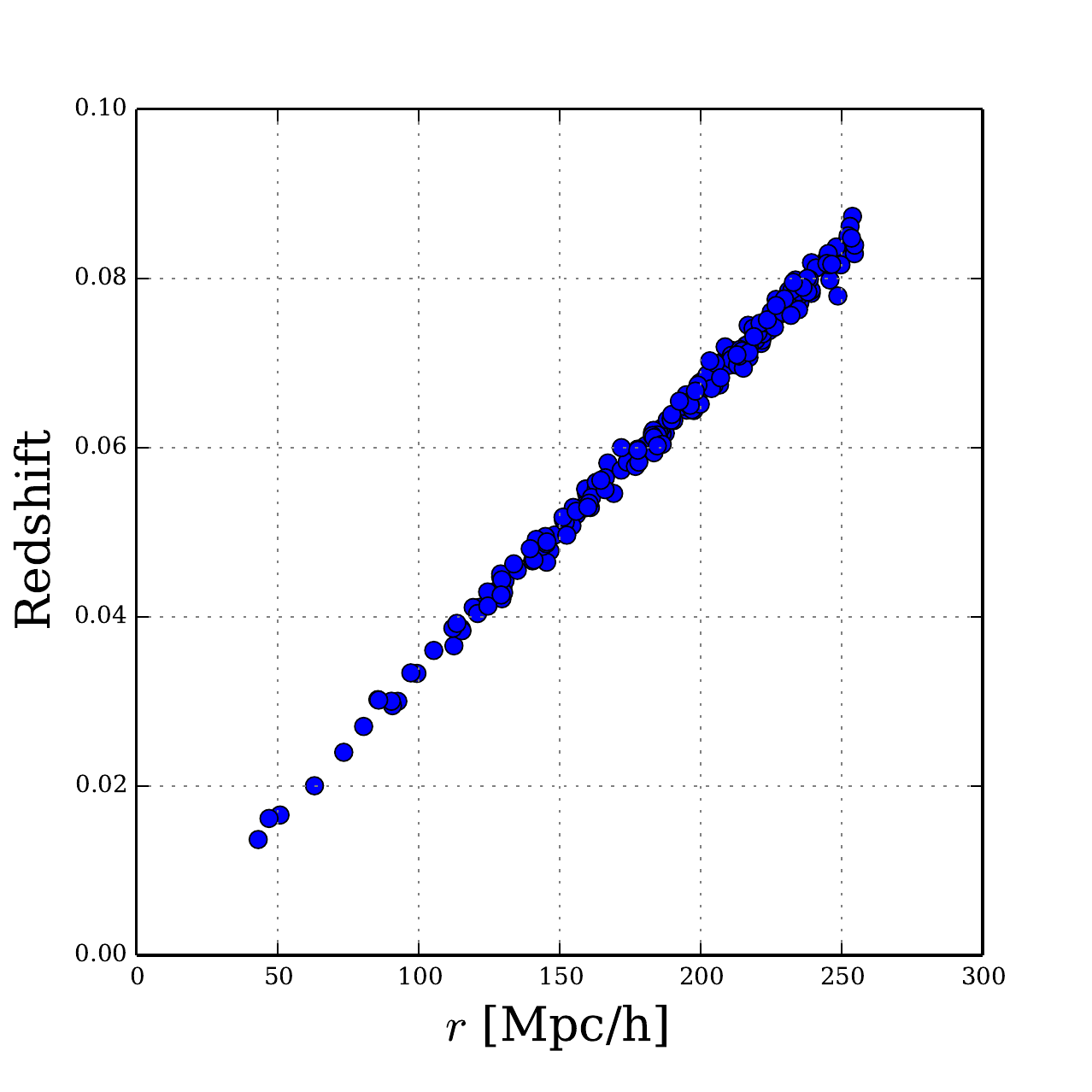}
\caption{Left: Comparison between the redshift distribution measured by a typical observer in the mock catalogues (blue) and the redshift distribution of supernovae from Hicken et al., \cite{Hicken:2009a} (green), which closely resembles the one used by Riess et al. in \cite{Riess:2011}. Right: Hubble diagram obtained by the same mock observer.} 
\label{fig:redshiftdistribution}
\end{figure}

In actual observations, the position and velocity information of distant halos is primarily obtained from supernovae. In \citep{Riess:2011}, 240 supernovae are used in estimating the local Hubble parameter, and for this reason we choose 240 observed halos for each of the observers in our mock survey. This is done by assuming that the probability of a supernova occurring in a given halo is proportional to the halo mass, and therefore making a mass-weighted selection of observed halos (this is in contrast to what is done in \cite{Wojtak:2013}, where every halo within a given distance is used). This procedure results in a redshift distribution of the observed halos which is peaked at a higher value than the one used in \cite{Riess:2011}, as shown in figure \ref{fig:redshiftdistribution}. In consequence we expect to slightly underestimate the variance of the local Hubble parameter at large distances.

There are some indications \cite{Perrett:2012,Rodney:2014} that the rate of type Ia supernovae not only depends on the stellar mass of a galaxy -- or more precisely the star formation history -- but there also exist a prompt component in the type Ia distribution correlated with the instantaneous star formation rate of the galaxy. Given the current uncertainty in the fraction of prompt type Ia supernovae, this has not been taken into account in the selection of halos. Instead we assume a direct proportionality between the type Ia supernovae rate, stellar mass, and halo mass.

When 240 observed halos have been picked for a given observer, they are distributed into bins according to their distance, with each bin characterized by the maximum distance of the observed halos it contains. The bin distances, which will henceforth be denoted by $r_{max}$, ranges from $67\di{Mpc/h}$ to $256\di{Mpc/h}$. The local Hubble constants are then estimated first using only the innermost bin, then the two innermost bins and so forth, until all the bins are included.
 
\subsection{Objectives}

\begin{figure}
\center
\includegraphics{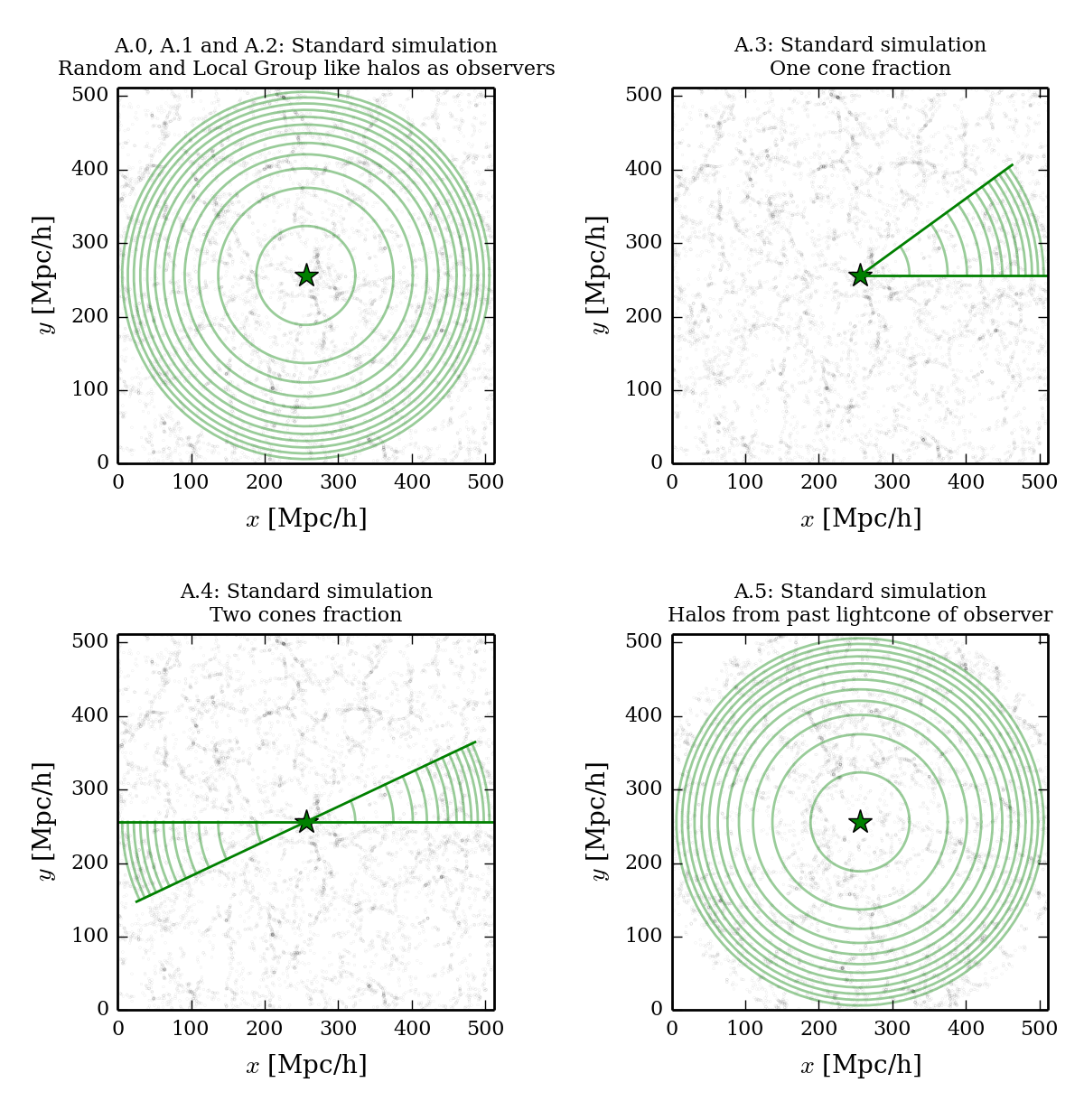} 
\caption{Positions of halos in a slap of thickness $10\di{Mpc/h}$, with illustrations of the  observation performed by each observer (green star), with values of $r_{max}$ shown as green circles (or arcs). For clarity, only every 5th value of $r_{max}$ is shown.  } 
\label{fig:refsheet}
\end{figure}

In order to determine how the variance of the local Hubble constants depend on various aspects of the observations, the observed halos are selected and handled in different manners, as is the choice of observer positions. 
Furthermore, effects of simulation seed, resolution and volume are checked for, and we additionally study how the number of observed halos affect the variance. 
Below we describe each of these analyses in more detail. In figure \ref{fig:refsheet} a few of the analyses are sketched for visualisation, and an overview is given in table \ref{tab:parameters}.

\begin{table}
\center
\begin{tabular}{cccccc}
Name & Nsim & Box [Mpc/h] & Survey geometry & Observers & Past lightcones \\ 
\hline 
A.0 & 512 & 512 & Full sky & Random in space & No \\ 
A.1 & 512 & 512 & Full sky & Random in halos & No \\
A.2 & 512 & 512 & Full sky & Local Group like & No \\ 
A.3 & 512 & 512 & One cone & Local Group like & No \\ 
A.4 & 512 & 512 & Two cones & Local Group like & No \\ 
A.5 & 512 & 512 & Full sky & Local Group like & Yes \\ 
B & 512 & 512 & Full sky & Local Group like & No \\
C & 1024 & 512 & Full sky & Local Group like & No \\ 
D & 1024 & 1024 & Full sky & Local Group like & No \\ 
\end{tabular}
\caption{Four different simulations have been run, named A-D. The number of particles in each simulation is Nsim$^3$. For the standard simulation (A), six different analyses have been performed, named A.0-A.5. Simulation B has the same characteristics as A, but with a different seed for the random number generator. All simulations are performed with the best fit parameters from the Planck satellite: $(\Omega_b,\Omega_{CDM},h,\sigma_8)=(0.048,0.26,0.68,0.84)$.} 
\label{tab:parameters}
\end{table}

\paragraph{Choice of observers:} 
The standard observer is chosen from the halo catalogues as a subhalo of mass $10^{12}-10^{13} M_{\odot}/\di{h}$ in a host halo with mass $5\cdot 10^{14}-5\cdot 10^{15} M_{\odot}/\di{h}$, since this approximately corresponds to our position in the Local Group galaxy cluster that resides in the Virgo Super Cluster. The significance of the observer positions is checked by using two alternative selections: One in which the observer positions consist of positions chosen randomly in the simulation volume, and one in which the observer positions are chosen randomly among all the halos in the simulation.  

\paragraph{Survey geometry:}

The actual supernovae used in estimating the expansion rate of the local universe are not distributed across the entire sky. In some directions, our sight is blocked, for example in the plane of the Milky Way. To account for this in the analysis, at first only a small patch of the sky is observed. This patch is subsequently enlarged until it covers all of the sky. The number of supernovae is the same no matter if the whole sky or only a fraction is observed. We perform this analysis both with one cone, and with two cones pointing in opposite directions.

\paragraph{Cosmic evolution:} 
Observing out to a redshift of $z\sim 0.1$, corresponds to observing 1.3 billion years back in time. When doing the same "observation" in the output of our N-body simulations, the time it takes light to travel from the distant halos is ignored, and therefore so are the changes in the growth and structure of the halos. We investigate the significance of this effect by using a locally developed plugin to GADGET-2 that during the course of the simulation reconstructs the past lightcone of any observer in the box, dumping it to a snapshot on disk. Such snapshots are created for the positions of our chosen observers. Then the halo finder is applied to these past lightcone snapshots, using a functionality in ROCKSTAR in which the change in cosmological parameters as one looks backwards in time is taken into account in the analysis. Constructing the mock observations using past lightcones takes into account the evolution in the cosmological model, and more accurately reproduces the actual observations.

\FloatBarrier
\section{Results}

\begin{figure}
\center
\includegraphics[width = 0.49\textwidth]{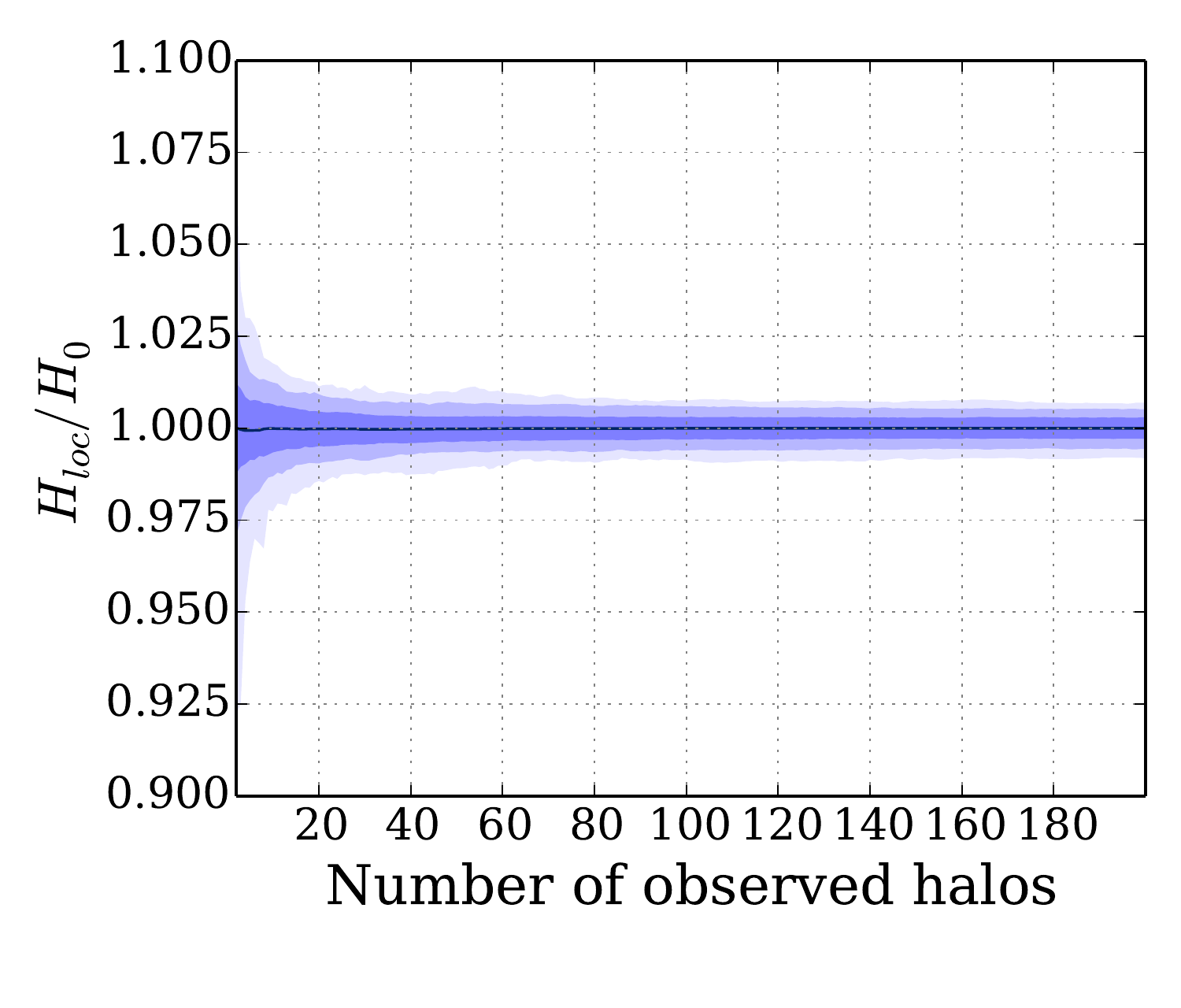} 
\includegraphics[width = 0.49\textwidth]{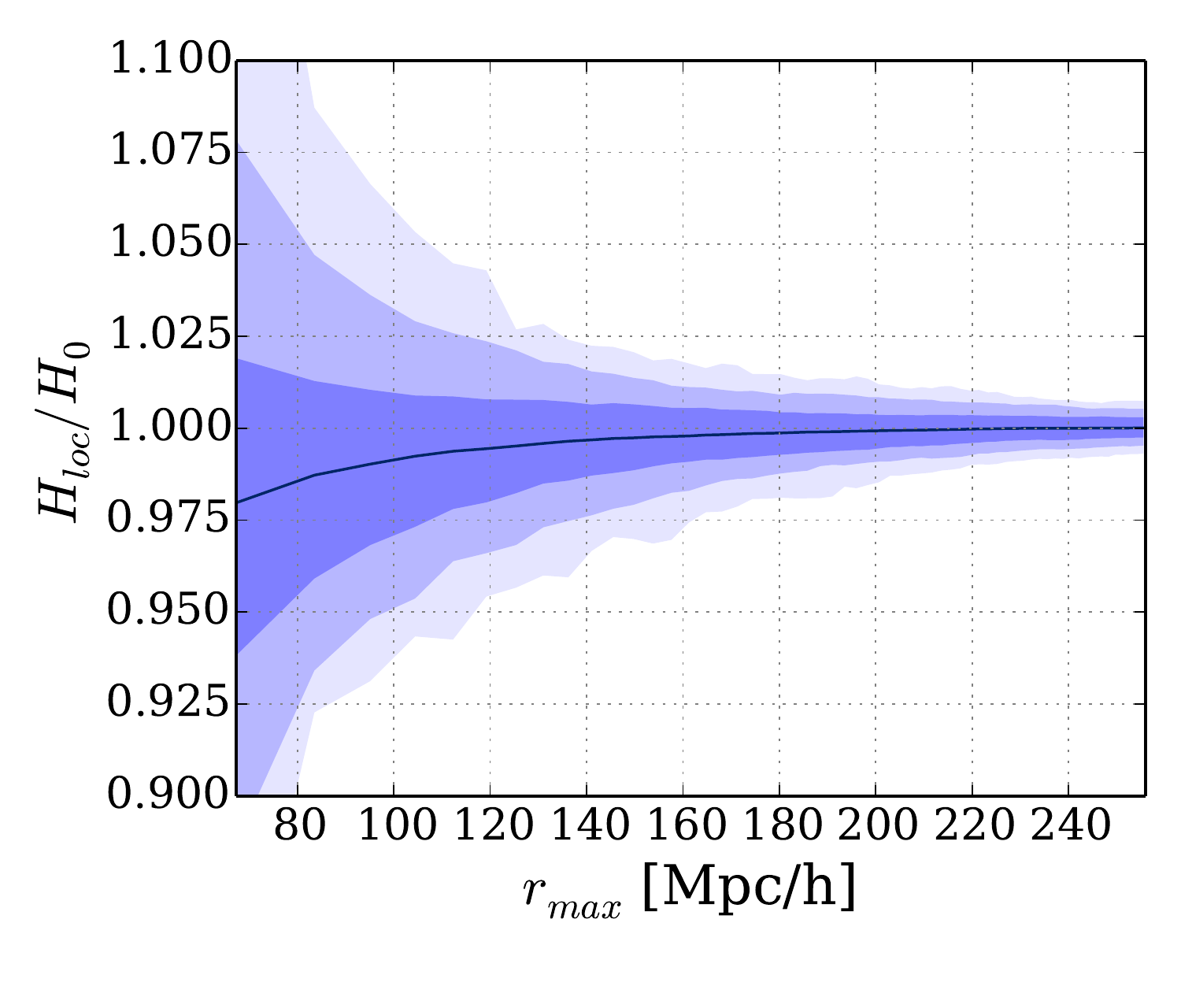}
\caption{Left: Mean and scatter of the local Hubble parameter as a function of the number of observed halos. The largest observed distance is held constant at $256\di{Mpc/h}$, while the number of observed halos is varied from 2 to 200. Right: Mean and scatter of the local Hubble parameter as a function of the maximum observed distance. The chosen observers are Local Group like halos, each estimating the local Hubble constant by measuring the distances and radial velocities of 240 supernovae in nearby halos. The simulation consists of $512^3$ particles in a periodic box of sidelength $512\di{Mpc/h}$.} 
\label{fig:Planck512}
\label{fig:Planck512_varSN}
\end{figure}

\begin{figure}[htb!]
\center
\includegraphics[width = 0.49\textwidth]{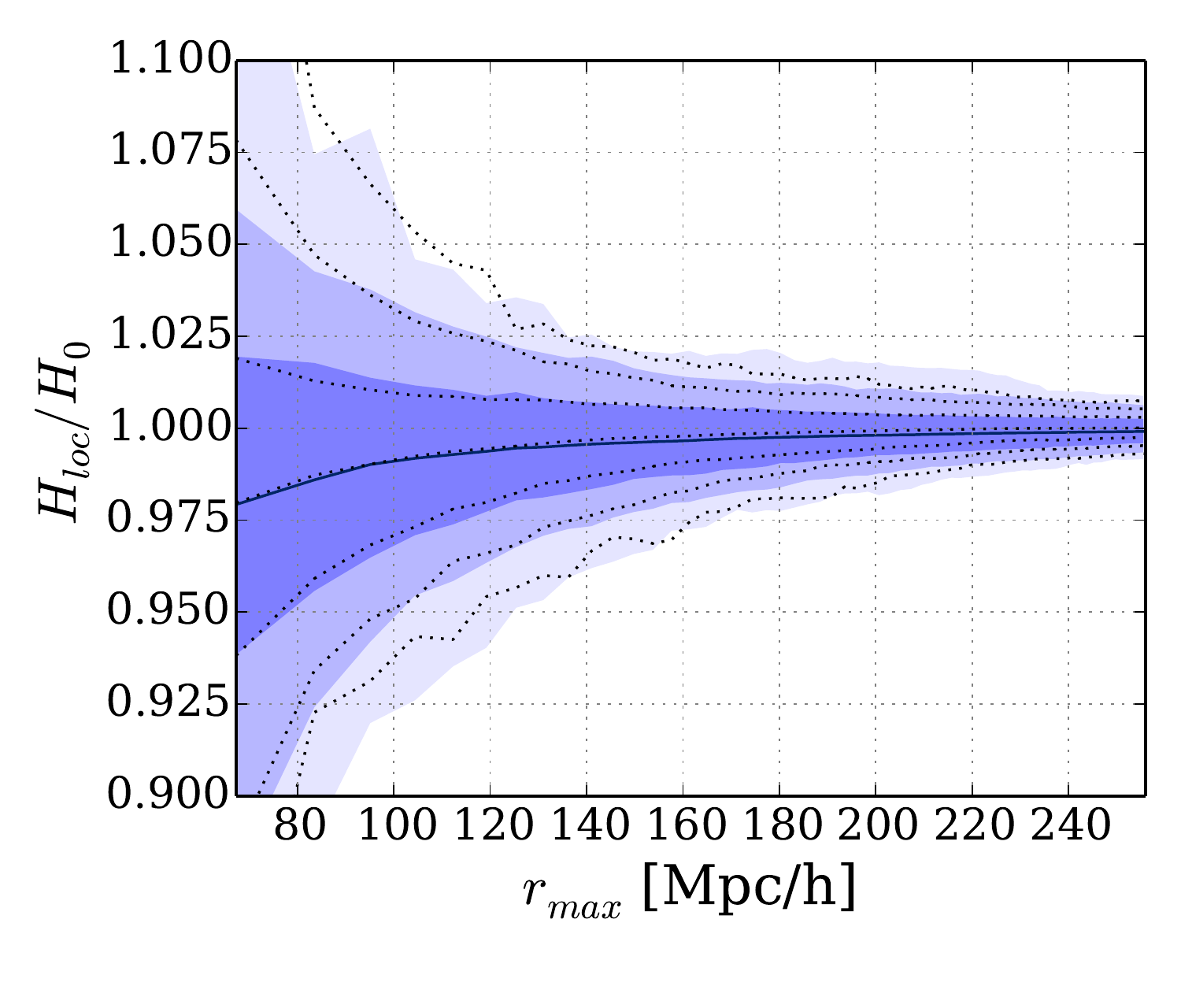} 
\includegraphics[width = 0.49\textwidth]{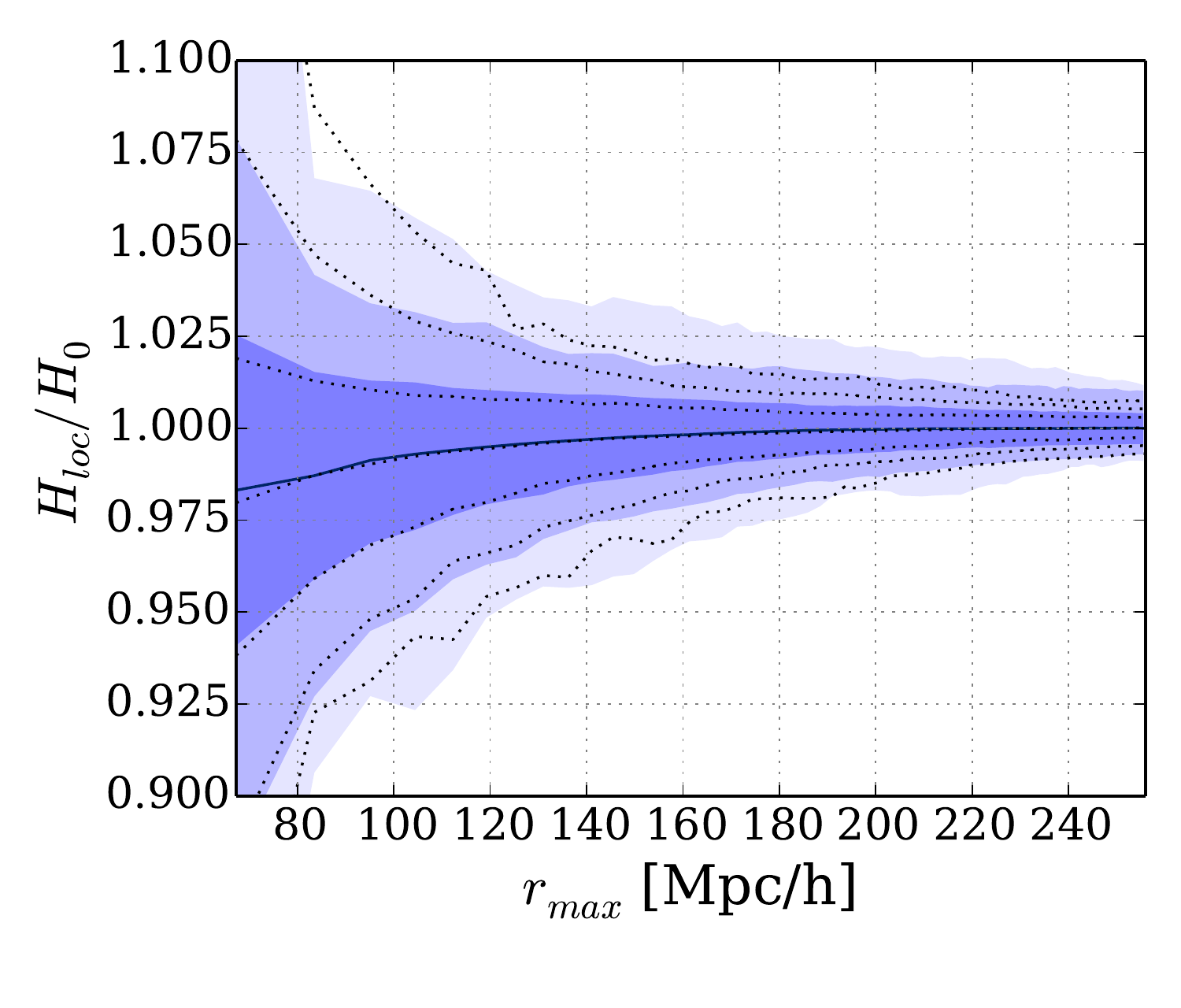} 
\caption{Left: Mean and scatter of the local Hubble parameter as a function of the maximum observed distance, calculated for a simulation with $1024^3$ particles in a box of $512 \di{Mpc/h}$. Right: Calculated for a simulation with $1024^3$ particles in a box of $1024 \di{Mpc/h}$. Punctuated lines indicate the result from the standard analysis.} 
\label{fig:Planck1024}
\label{fig:Planck1024_box1024}
\end{figure}

\begin{figure}[htb!]
\center
\includegraphics[width = 0.49\textwidth]{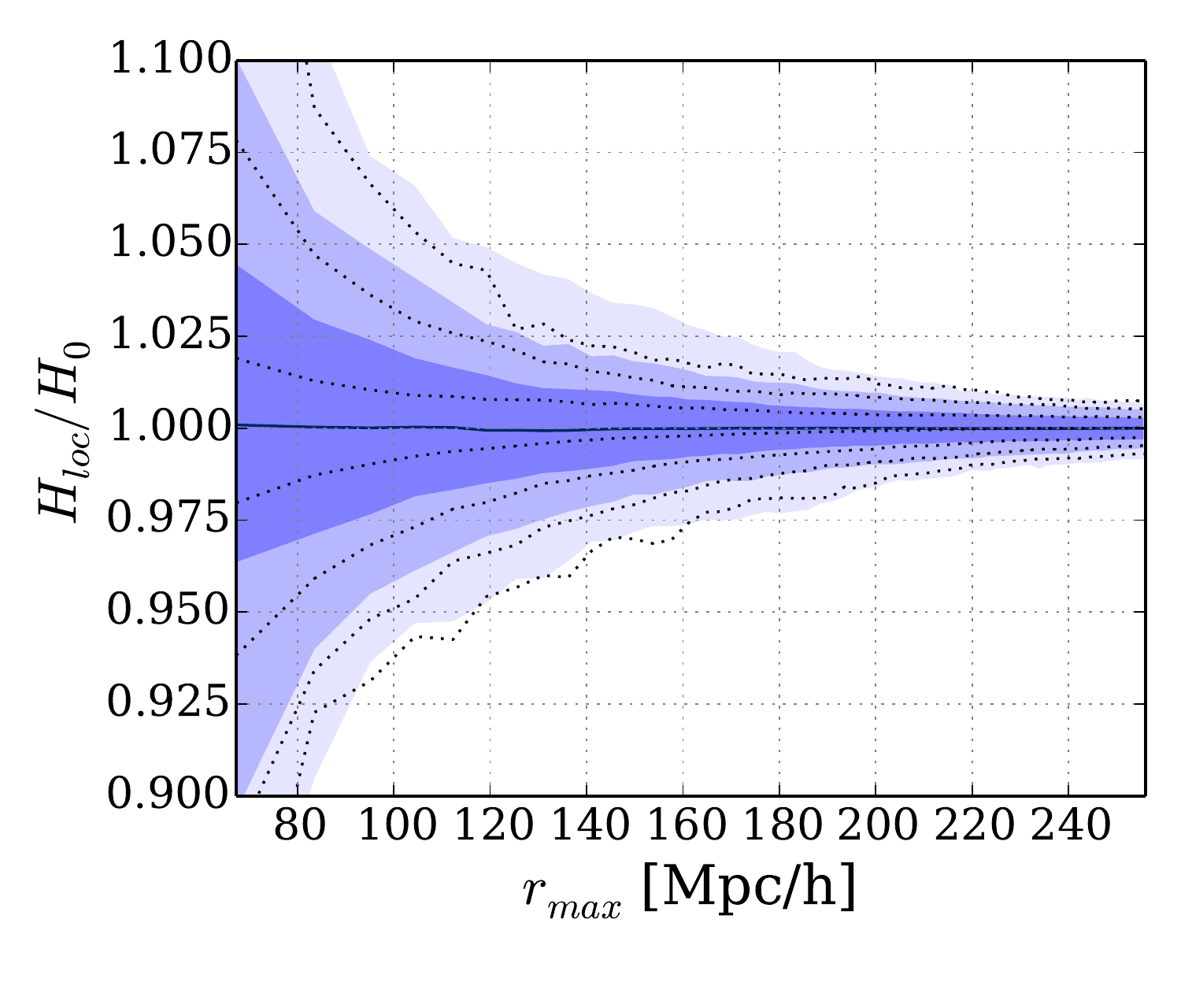} 
\includegraphics[width = 0.49\textwidth]{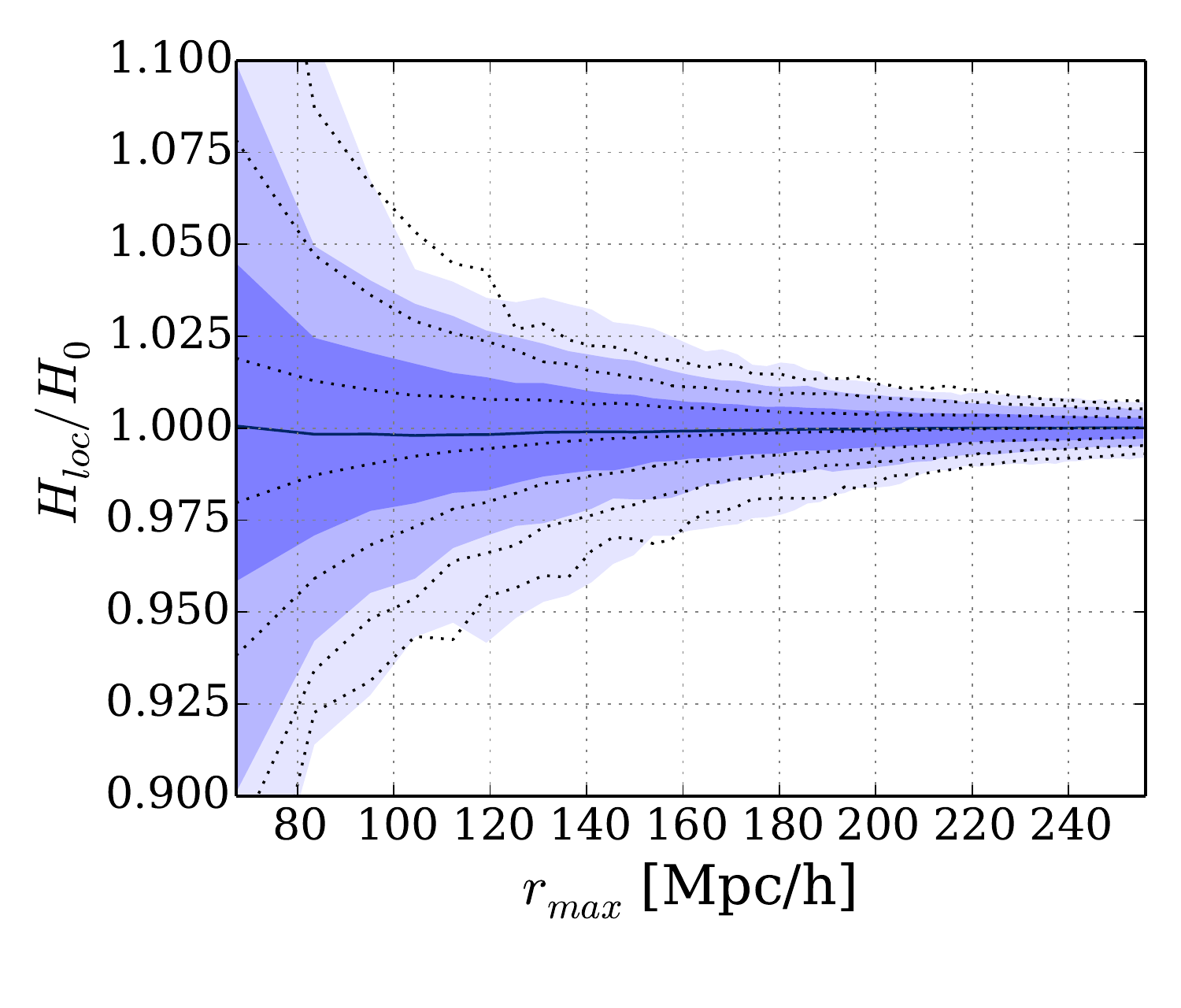} 
\caption{Left: Mean and scatter of the local Hubble parameter as a function of the maximum observed distance, when the chosen observers are distributed randomly in the simulation volume. Right: Observers chosen randomly among all the halos in the simulation. Punctuated lines indicate the result from the standard analysis.} 
\label{fig:Planck512_RandPos}
\label{fig:Planck512_RandHalos}
\end{figure}

\begin{figure}[htb!]
\center
\includegraphics[width = \textwidth]{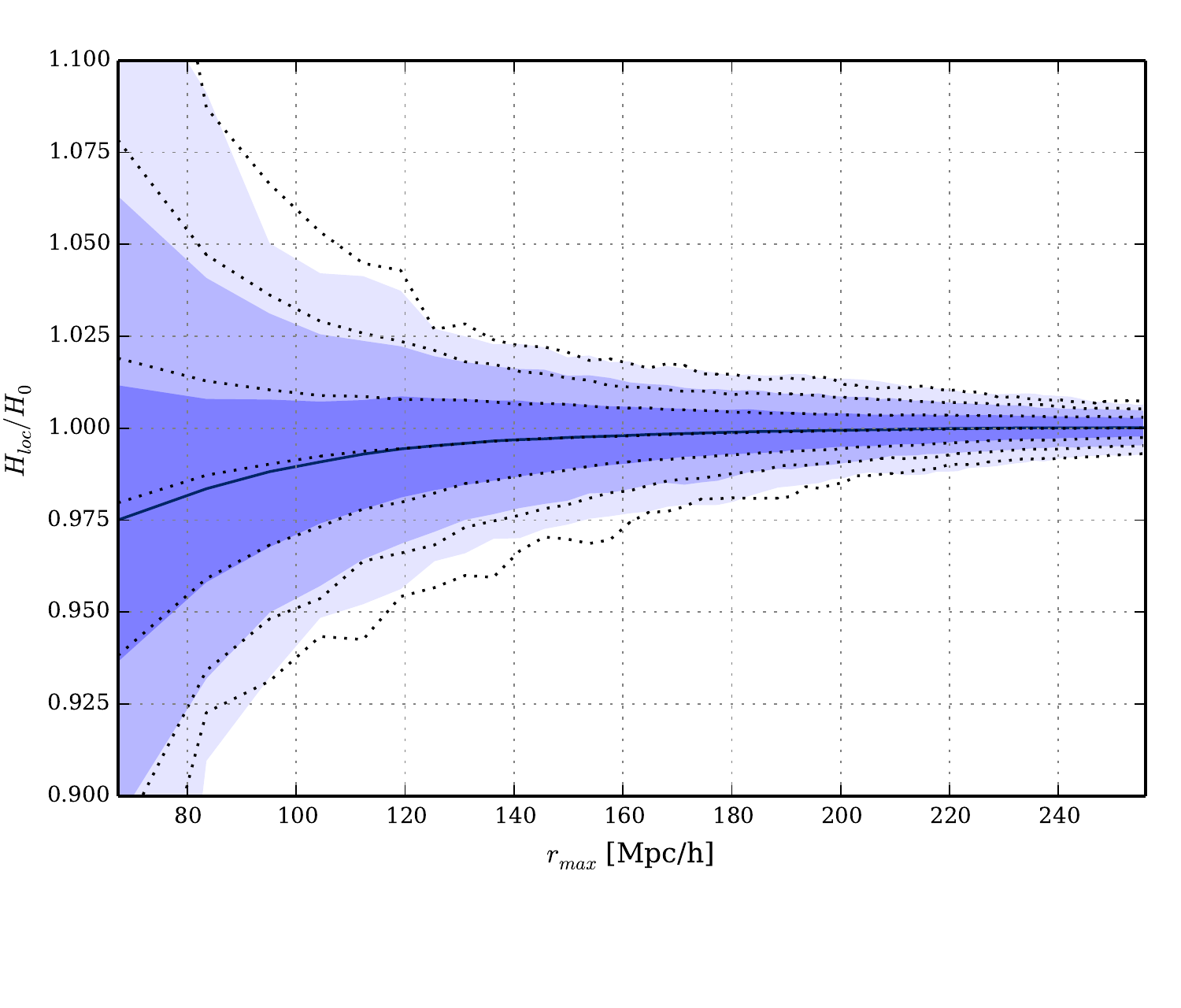} 
\caption{Mean and scatter of the local Hubble parameter as a function of the maximum observed distance calculated for Local Group like halos, with the effect of cosmic evolution taken into account by using lightcones. Punctuated lines indicate the result from the standard analysis.} 
\label{fig:Planck512_lightcone}
\end{figure}

In Figures \ref{fig:Planck512}-\ref{fig:Planck512_lightcone} we show how the mean of $H_{loc}/H_0$ among the chosen observers depend on the number of observed halos and the maximum observed distances for different choices of observers and observed halos. Confidence intervals are shown as coloured bands around the mean (solid line). The chosen confidence levels are 68.3, 95.4 and 99.7 per cent, all calculated symmetrically so that equal fractions of the locally measured Hubble constants fall below and above the interval. For comparison, the results from the standard analysis (on the right in figure \ref{fig:Planck512}) are shown as punctuated lines in all other plots.\\
\\
\noindent
The first plot in figure \ref{fig:Planck512_varSN} shows $H_{loc}/H_0$ as a function of the number of observed halos. In figure \ref{fig:Planck1024}, the analysis is repeated for two simulations with $1024^3$ particles, in boxes of sidelength $512\di{Mpc/h}$ and $1024\di{Mpc/h}$, respectively. This is done order to check that the simulation resolution and simulation volume have no significant effect on the results.
The set of parameters used in the N-body simulations are the ones published by the Planck collaboration:
$(\Omega_b,\Omega_{CDM},h,\sigma_8)=(0.048,0.26,0.68,0.84)$. 

\begin{figure}[htb!]
\center
\includegraphics[width = \textwidth]{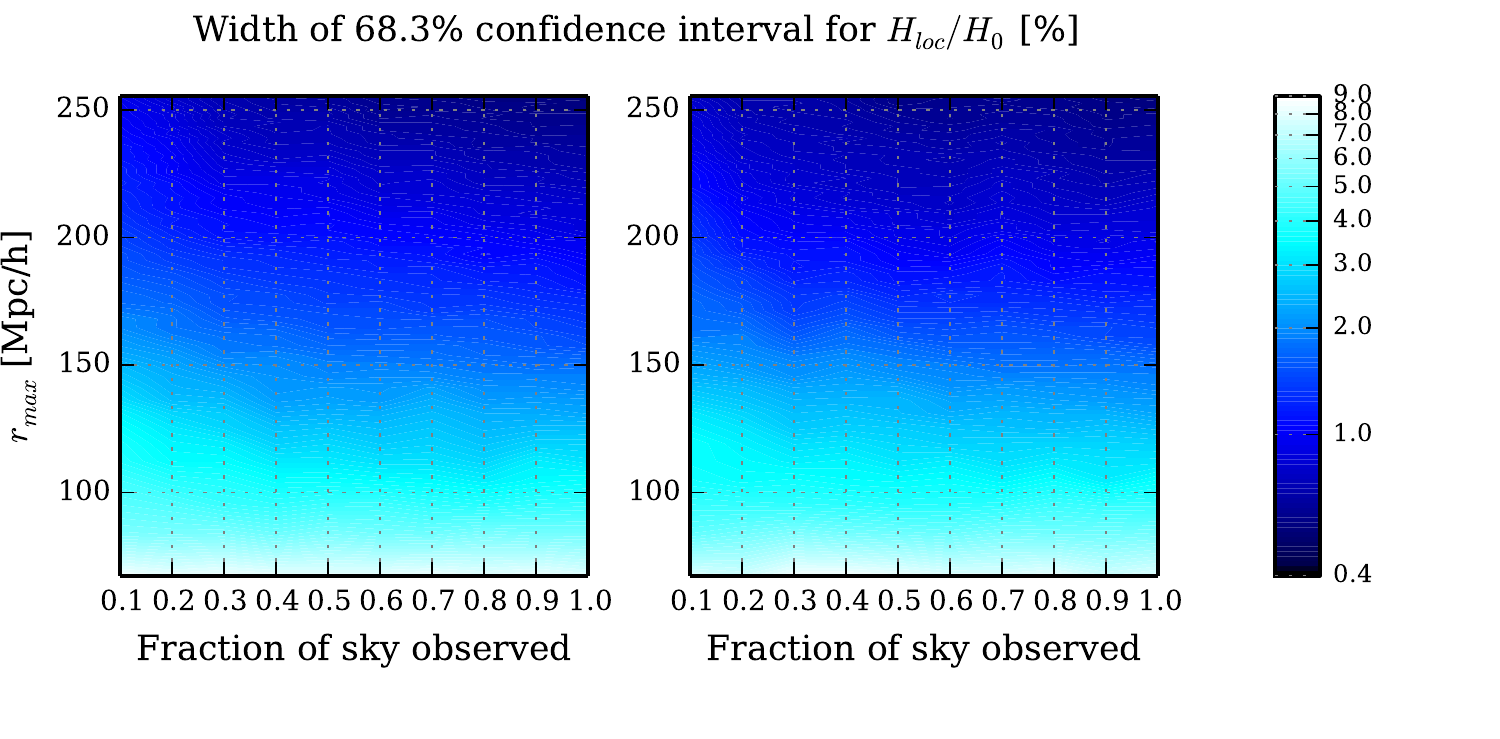} 
\caption{Width of 68.3\% confidence interval for $H_{loc}/H_0$ when observing only a fraction of the sky. Left: One cone. Right: Two cones, pointing in opposite directions of the sky.} 
\label{fig:Planck512_skyfraction}
\end{figure}
\begin{figure}[htb!]
\center
\includegraphics[width = \textwidth]{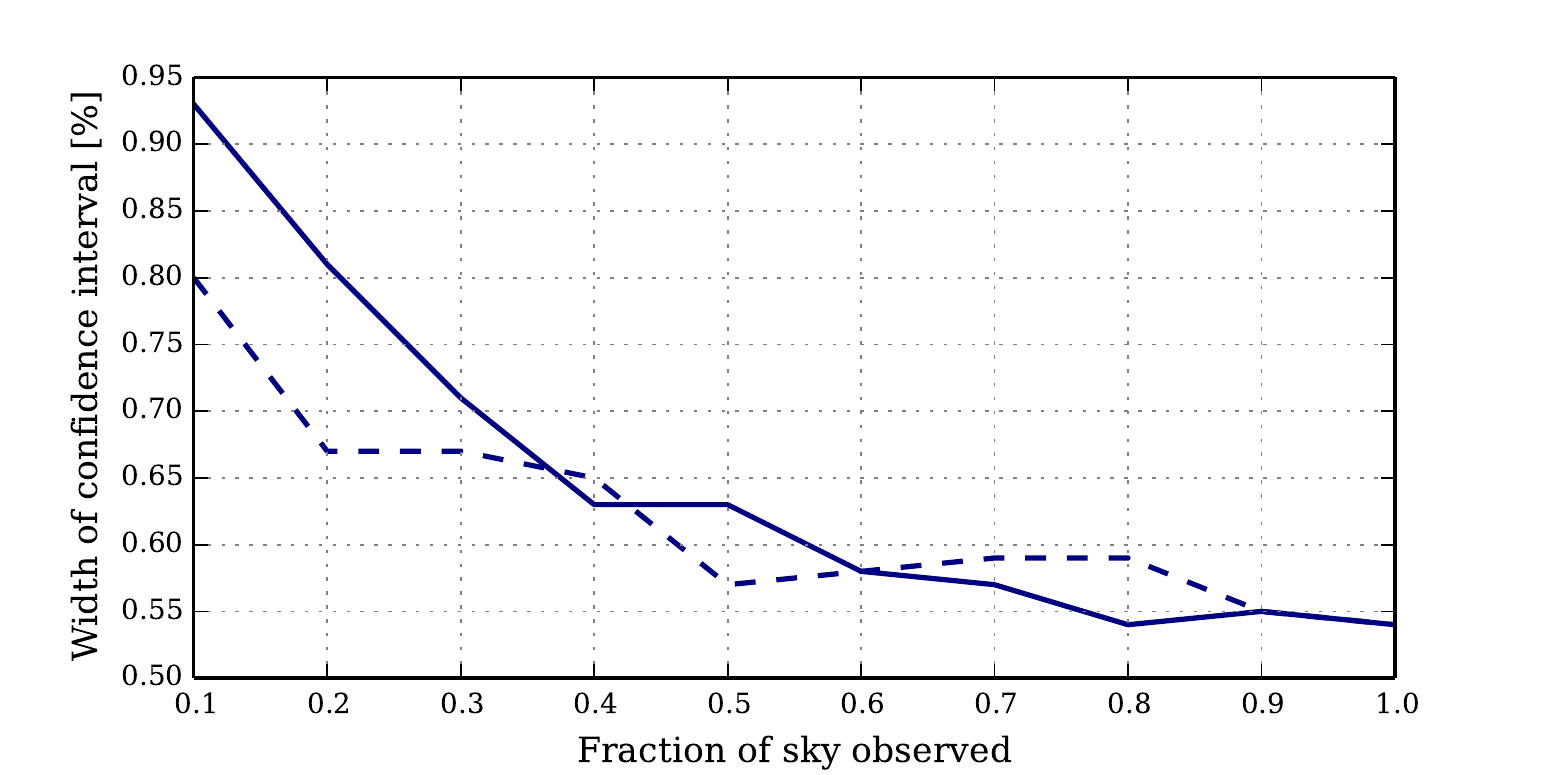} 
\caption{Width of 68.3\% confidence interval for $H_{loc}/H_0$ as a function of the observed fraction of the sky at $r_{max}=256\di{Mpc/h}$. The full line corresponds to observations in one cone, the punctuated line to observations in two cones pointing in opposite directions.} 
\label{fig:Planck512_skyfraction_256}
\end{figure}

In figure \ref{fig:Planck512_RandPos} we show how the observer positions affect the measured values of $H_{loc}$ (on the left). On the right, the observers are chosen randomly among all the halos in the simulation. In figure \ref{fig:Planck512_lightcone}, the observations are performed using past lightcones of the Local Group like observers, so that cosmic evolution from the time that light was emitted from an observed halo is taken into account. In figure \ref{fig:Planck512_skyfraction} we show how the width of the 68.3 per cent confidence interval varies as a function of both the distance $r_{max}$ and the covered percentage of the sky, and in figure \ref{fig:Planck512_skyfraction_256} the width of the confidence interval is plottet as a function of the observed skyfraction at $r_{max}=256\di{Mpc/h}$. The results are summarized in table \ref{tab:results}.

\begin{table}
\center
\begin{tabular}{cccccccc}
Name & Description & $\mu_{67}[\%]$ & $\mu_{150}[\%]$ & $\mu_{256}[\%]$ & $\sigma_{67}[\%]$ &  $\sigma_{150}[\%]$ &  $\sigma_{256}[\%]$ \\ 
\hline 
• & \textbf{Observers} &  • & • & • & • & • & • \\ 
A.0 & Random positions in space & 0.1 & 0.0 & 0.0 & 4.6 & 0.9 & 0.3 \\
A.1 & Random positions in halos & 0.1 & -0.1 & 0.0 & 4.7 & 0.9 & 0.3 \\
A.2 & Local Group-like halos & -2.0 & -0.3 & 0.0 & 4.5 & 0.9 & 0.2 \\
\hline
• & \textbf{10\% sky coverage}&  • & • & • & • & • & • \\
A.3 & One cone &  -2.6 & -0.2 & -0.0 & 4.8 & 1.2 & 0.4 \\
A.4 & Two cones & -2.6 & -0.3 & -0.0 & 4.2 & 1.1 & 0.4 \\
\hline
• & \textbf{Cosmic evolution}& • & • & • & • & • & • \\
A.5 & Lightcone & -2.5 & -0.3 & 0.0 & 4.4 & 0.9 & 0.2 \\
\hline
• & \textbf{Simulation}& • & • & • & • & • & • \\
B & Different seed & -1.7 & -0.2 & 0.0 & 4.6 & 0.9 & 0.4 \\
C & Box=512, Nsim=1024 &  -2.1 & -0.4 & -0.1 & 4.6 & 1.0 & 0.3 \\
D & Box=1024, Nsim=1024 &  -1.7 & -0.2 & 0.0 & 5.0 & 1.1 & 0.4 \\
\end{tabular}
\caption{Mean ($\mu$) and variance ($\sigma$) of $H_{loc}/H_0$ at selected distances. The subscripts specify the distance in $\di{Mpc}/h$. All results are given in per cent.}
\label{tab:results}
\end{table}

\FloatBarrier
\section{Discussion}

\begin{figure}[htb!]
\center
\includegraphics[width = \textwidth]{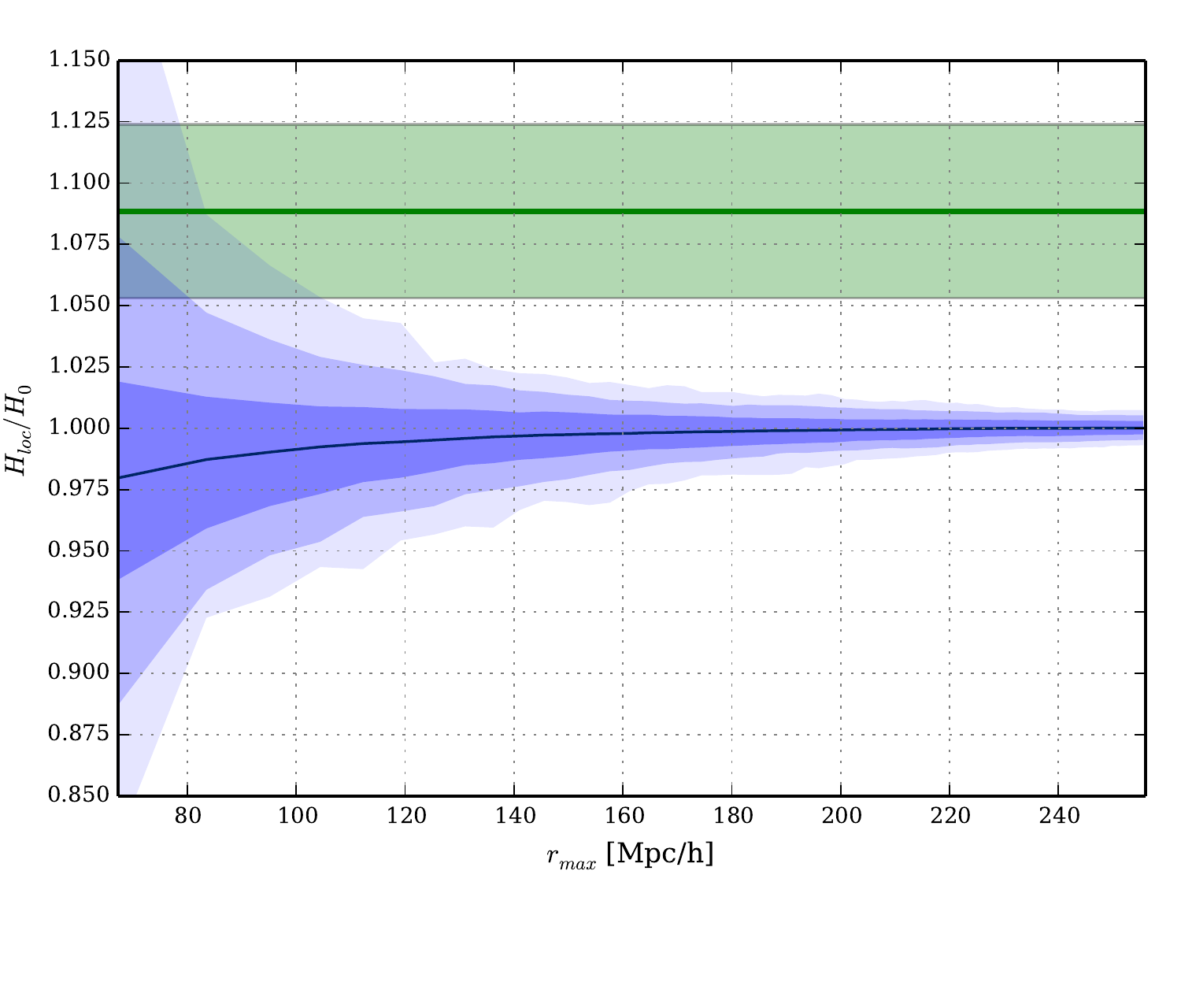} 
\caption{The value of the local Hubble parameter found by Riess et al. (green line), is plotted together with the value and spread of $H_{loc}/H_0$ found in the standard analysis. The green band indicates the $1\sigma$ uncertainty in the result from Riess et al.} 
\label{fig:Planck512_band}
\end{figure} 

\noindent
Figure \ref{fig:Planck512_band} shows the mean and variance of the standard simulation (as in figure \ref{fig:Planck512}) together with the value of the Hubble parameter found by Riess et al. in \cite{Riess:2011} of $73.8\pm 2.4$, plotted as a green line, with the uncertainty of the measured value shown as a green band. Only for very small values of $r_{max}$ is there any overlap between the measured range of $H_0$ and the confidence bands obtained from a Planck like universe. At a distance of $r_{max} = 256\di{Mpc}/h$, less than $0.3\%$ of the Local Group like observers would observe a value as high as the one we see. We therefore conclude that the variance of the expansion field does not lift the discrepancy between the Hubble constant determined from measurements of the CMB and that obtained by direct measurements of recessional velocities in our local universe. This has also been concluded by Marra et al. in \cite{Marra:2013} and by Wojtak et al. in \cite{Wojtak:2013}. At an intermediate distance of $150\di{Mpc/h}$, we find the width of the 63.8\% confidence interval for the local Hubble constant to be $0.9\%-1.1\%$ (depending on the simulation), which is in good agreement with the value of $0.9\%$ found by Wojtak et al. Using equation 5 from \cite{Marra:2013} and the redshift distribution from \cite{Hicken:2009a}, we compute the standard deviation of $H_{loc}/H$ in the redshift range $0.023 < z < 0.1$. Averaging over all observers, we get 1.1\% for the standard simulation. This is 38\% bigger than the value of 0.8\% found by Wojtak et al., and 8\% smaller than the value of 1.2\% found by Marra et al. from analytical calculations based on the power spectrum.    
We observe a tendency for the local Hubble parameter as measured by observers in halos to be systematically lower than the overall expansion rate, whereas observers distributed randomly in space tends to overestimate $H_0$. This can be explained by noting that observers in halos are positioned in in-fall regions as a consequence of ongoing structure formation, whereas observers positioned at random will have a tendency to be located in regions less dense than average, because these take up a greater fraction of the simulation volume than the overdense regions. Both effects are smoothed out when observing halos at large distances. This is in good agreement with the result obtained in \cite{Wojtak:2013}. Observers positioned randomly in halos will in mean observe a Hubble constant very close the actual value.

\section{Conclusion}

We have carefully studied how local measurements of $H_0$ can be influenced by a variety of different parameters related to survey geometry, depth, and size, as well as observer position in space.

We find that variations in the local expansion field {\it cannot} explain the difference in the Hubble parameter obtained indirectly by measurements of the CMB by the Planck collaboration and that obtained by direct measurements. This result has been found to be insensitive to the percentage of the sky observed in direct measurements, and to whether or not cosmic evolution is taken into consideration. At small distances, observers positioned in Local Group like halos will have a tendency to measure a Hubble constant that is lower than the true value, whereas observers positioned at random in space have a tendency to measure a higher value. However, these effects become negligible when the largest observed distance exceed a few hundred $\di{Mpc/h}$, and therefore have no significant effect on the scale used in the local measurement of the Hubble parameter. 

Our conclusion is that the discrepancy between the value of $H_0$ inferred from Planck and the value found from direct measurements must be ascribed to other sources than a variation in the local velocity field.

\acknowledgments
We acknowledge computing resources from Center for Scientific Computing Aarhus. Research at Centre for Star and Planet Formation is funded by the Danish National Research Foundation. TH is supported by a Sapere Aude Starting Grant from The Danish Council for Independent Research. 
\bibliographystyle{utcaps}
%
%
\providecommand{\href}[2]{#2}\begingroup\raggedright\endgroup

\end{document}